\documentstyle[preprint,revtex]{aps}
\begin {document}
\draft
\preprint{UCI-TR 92-56/Uppsala U. isv PT 14 1992 }
\begin{title}
On the 545 keV
Line in the Spectrum of the Crab Nebula
\end{title}
\author{Myron Bander\footnotemark\ }
\begin{instit}
Department of Physics, University of California, Irvine, California
92717, USA
\end{instit}
\author{H. R. Rubinstein\footnotemark\ }
\addtocounter{footnote}{-1}\footnotetext{e-mail:
mbander@funth.ps.uci.edu}\addtocounter{footnote}{1}%
\footnotetext{e-mail:
rub@vand.physto.se}
\begin{instit}
Department of Radiation Sciences, University of Uppsala, Uppsala,
Sweden
\end{instit}
%\receipt{March\ \ \ 1991}
\begin{abstract}

\end{abstract}

%\narrowtext
The GRANAT group (R. Sunyaev {\it et al.\/}, Central Bureau of Astronomical
Telegrams, International Astronomical Union, Circular No.~5481) recently
reported the observation of a $(545  \pm 11)$ keV line in the spectrum of
the Crab Nebula. It is tempting to associate this with the
positron-electron annihilation line at 511 keV. If this line originates
from some transition at the surface of the neutron star in the Crab, we
expect a red shift of around 10\% rather then the observed blue shift of
$\sim 35$ keV; in fact, this gravitational red shift implies that the
natural frequency of the observed line is also 10\% higher, or around 600
keV. In this article we shall present a mechanism for generating such a
split from 511 keV. For this mechanism to be successful we require: (a)
surface magnetic fields, in the annihilation region, to be $(3-8)\times
10^{12}$~G and to point in our direction, (b) surface densities of
$(10^5-10^6)$ g/cc. Both of these assumptions are consistent with pulsar
models.

Models of pulsating neutron stars \cite{Ruderman,Alpar} require surface
magnetic fields
$B=(10^{12}-10^{13})$ G. Electrons in such fields arrange themselves into
degenerate Landau levels \cite{IZ} with quantum numbers $n,m$ ($m$ is
the orbital angular momentum along the magnetic field) and energies
\begin{equation}
E_{n,m,s}=\sqrt{M^2+(2n+1+s)|eB|}\, .\label{Landauenergy}
\end{equation}
$M$ is the mass of the electron and $s=\pm 1$ depending on whether the
electron spin is along or opposite to the magnetic field. In the above, and
subsequently, we shall ignore motion along the field lines. The energies
of positrons in such a field are obtained from Eq.~(\ref{Landauenergy}) by
letting $s\rightarrow -s$ in the right hand side of that equation. For
electrons the lowest energy state is the one with $n=0,\ s=-1$ while the
next highest energy state has quantum numbers $n=0,\ s=+1$ or $n=1,\ s=-1$;
\begin{eqnarray}
E_{0,m,-1}&=&M\nonumber\\
E_{0,m,+1}&=&\sqrt{M^2+2|eB|}\nonumber\\
E_{1,m,-1}&=&\sqrt{M^2+2|eB|}\, .\label{lowestlelels}
\end{eqnarray}
The energies of the lowest positron levels are obtained by changing the
signs of the spins. We propose that the observed line is due to the
annihilation of a positron in the $(0,m,+1)$ state with an electron in
 the $(0,m,+1)$ state. (It will turn out that only
states with $m=0$ or $m=1$ contribute significantly to this annihilation.)
For this mechanism to be valid we require magnetic fields $B=3\times
10^{12}$ G, for no gravitational red shift or $B=8.4\times 10^{12}$ G for
a 10\% red shift. Consistent with magnitudes postulated to exist
on the surface of pulsating neutron stars.

We take the positrons to be in the ground state and their
density is assumed not to be high enough to fill more than
one Landau level; any positrons in
higher energy levels will very rapidly decay into the ground state. How
many levels the electrons fill up depends on the density. As no peak
corresponding to annihilation with electrons in levels with energies
$\sqrt{M^2+2|eB|}$ are seen this level must be empty. This forces the
electron Fermi level, $\epsilon_F$, to satisfy $M <\epsilon_F\le
\sqrt{M^2+2|eB|}$ and in turn the density, $\rho$, in units of
g/cc, to be bounded by $10^5\le\rho\le 3\times 10^5$ for the case of no
gravitational red shift and $5\times 10^5\le\rho\le 1.5\times 10^6$
for a 10\% red shift. These are densities conjectured to exist on the
magnetic surface next to the outer crust of a neutron star. It still
remains to be shown why no
annihilation with electrons in the lowest state is observed. To answer
this question we must examine some of the details of the annihilation process.

To lowest order in $eB/M^2$ the annihilation amplitude is \cite{IZ2}
\begin{equation}
A(n_p,m_p,s_p;n_e,m_e,s_e)=\int d^2P_ed^2P_p\
u^{\dag}(s_p)\psi^{*}_{n_p,m_p}(P_p){\cal
M}(P_p,P_e)\psi_{n_e,m_e}(P_e)u(s_e)\, ; \label{amplitude}
\end{equation}
the $u$'s are two component spinors, $\psi_{n,m}(P)$ is the momentum space
wave function for the Landau level $(n,m)$ and
\begin{equation}
{\cal M}(P_p,P_e)={{\sigma_y}\over
{2M^2}}\Bigl[\sigma\cdot\epsilon_1\
(\Pi_e-\Pi_p)\cdot\epsilon_2+\sigma\cdot\epsilon_2\
(\Pi_e-\Pi_p)\cdot\epsilon_1\Bigr]\, ;\label{matrixelement}
\end{equation}
$\Pi_e=P_e-eA$, $\Pi_p=P_p+eA$, where $A$ is the vector potential
responsible for the magnetic field.  Looking at the rotation (around
the B axis) properties of the wave functions one notes that the terms
proportional to $\Pi_e$ connect a positron with $m=0$ to an electron
with $m=1$ and the terms involving $\Pi_p$ connect a positron with
$m=1$ to electrons with $m=0$. Thus in the annihilation process the
orbital angular momentum about the magnetic field changes by one unit,
$\Delta l=\pm 1$. For the electron and positron spins aligned the
change in the total angular momentum component along the magnetic
field, $j$, satisfies $\Delta j=0,\pm 2$ while for the spins opposite to
each other $\Delta j=\pm 1$. In the fist case both photons can emerge
along the field direction, while in the second there must be a photon
momentum component transverse to the field. Eq.~(\ref{amplitude}) and
Eq.~(\ref{matrixelement}) give explicit distributions consistent with
these arguments. The photon angular distributions are:
\begin{eqnarray}
d\Gamma(0,0,+1;0,1,-1)&=&2C\cos^2(\theta)\sin^2(\theta)d\cos(\theta)
\, ;\nonumber\\
d\Gamma(0,0,+1;0,1,+1)&=&C\cos^4(\theta)d\cos(\theta)\, ;
\label{distribution}
\end{eqnarray}
the first set of indices refer to the quantum numbers of the positron while
the second  to those of the electron.  $\theta$ is the angle of the emitted
photon with respect to the magnetic field while $C$ is a constant.
The distribution for an annihilation of positrons with $m=1$ and electrons
with $m=0$ has the same form. For directions along the magnetic field
photons from annihilations with electrons whose spin quantum number $s=+1$
will predominate (in a $30^\circ$ cone around the magnetic field the
intensity of this line will be a factor of 14 larger than that for the line
coming from annihilations with electrons in the lowest state). Thus if the
magnetic field responsible for the splitting points in our direction, lines
from annihilation of electrons and positrons with parallel spins will be
much stronger than those from antiparallel spins and we will not see
transitions with electrons in the ground state. As this annihilation
occurs at the outer crust it may be susceptible to surface glitches and
thus be episodic. Transitions between Landau levels should occur.
These are the same as the cyclotron radiation observed for other
systems \cite{Trumper}. For the present case these shoul be at $35$
keV for the no red shift case and $80$ keV for a 10\% red shift. A
simultaneous measurement of annnihilation lines and cyclotron lines
will determine both the magnetic field and $M_P/R_P$ of the pulsar.

\centerline{ACKNOWLEDGEMENTS}
We wish to thank Dr. Jean Swank for making us aware of the observations and
for many discussions. We also wish to thank Dr. G. Chanan and Dr. A.
Goobar for valuable
information. M.\ B.\ was  supported in part by the National Science
Foundation. H.\ R.\ was supported by the Swedish Research Council and
an EEC Science grant.

%\narrowtext

\newpage
\begin{references}
\bibitem{Ruderman}
Ruderman, M. {\it Ann. Rev. Astron. Astrophys.\/} {\bf 10}, 427-476
(1972).
\bibitem{Alpar}
Alpar, M.\ A. in {\it Timing Neutron Stars}, ed. H.\ {\"O}gelman and E.\
P.\ J.\ van den Heuvel, (Dordrecht, Kluver Academic Publishers, 1988),
p. 431.
\bibitem{IZ}
Itzykson, C.\ and Zuber, J.\-B.,, {\it Quantum Field Theory} (New York:
McGraw Hill, 1980), p. 67.
\bibitem{IZ2}
{\it ibid.\/}, p. 230.
\bibitem{Trumper}
Tr\"umper, J. {\it et al. Astrophys. J.\/} {\bf 219}, L105-L110 (1978).
\end{references}
\end{document}